\UseRawInputEncoding
\documentclass[12pt,3p]{article}

\usepackage{amssymb}
\usepackage{graphicx}
\usepackage{amsthm,amssymb,amsfonts,mathrsfs,amsmath}
\usepackage[]{inputenc}
\usepackage[english]{babel}

\usepackage{epstopdf}
\usepackage{algorithm}
\usepackage{algorithmic}
\usepackage{multicol,subcaption}
\usepackage{cite}
\usepackage{pdflscape}
\usepackage{array}
\usepackage{booktabs}
\usepackage{tabularx,ragged2e,booktabs}
\newtheorem{definition}{Definition}

\usepackage{lineno}
\makeatletter
\def\makeLineNumberLeft{%
  \linenumberfont\llap{\hb@xt@\linenumberwidth{\LineNumber\hss}\hskip\linenumbersep}
  \hskip\columnwidth
  \rlap{\hskip\linenumbersep\hb@xt@\linenumberwidth{\hss\LineNumber}}\hss}
\leftlinenumbers
\makeatother

\begin{document}

\title{Centrality Measures in Complex Networks: A Survey}


\author{
Akrati Saxena \\
  Department of Mathematics and Computer Science\\
  Eindhoven University of Technology, Netherlands\\
  \texttt{a.saxena@tue.nl}
  \and
 Sudarshan Iyengar \\
   Department of Computer Science and Engineering\\
  Indian Institute of Technology Ropar, India\\
  \texttt{sudarshan@iitrpr.ac.in} 
}
\date{}
\maketitle


\tableofcontents
\pagebreak

\begin{abstract}
In complex networks, each node has some unique characteristics that define the importance of the node based on the given application-specific context. These characteristics can be identified using various centrality metrics defined in the literature. Some of these centrality measures can be computed using local information of the node, such as degree centrality and semi-local centrality measure. Others use global information of the network like closeness centrality, betweenness centrality, eigenvector centrality, Katz centrality, PageRank, and so on. In this survey, we discuss these centrality measures and the state of the art literature that includes the extension of centrality measures to different types of networks, methods to update centrality values in dynamic networks, methods to identify top-k nodes, approximation algorithms, open research problems related to the domain, and so on. The paper is concluded with a discussion on application specific centrality measures that will help to choose a centrality measure based on the network type and application requirements.
\end{abstract}

\section{Introduction}
Complex networks \cite{kochen1989small, strogatz2001exploring} are encountered frequently in our day to day lives such as World Wide Web \cite{albert1999internet, broder2000graph}, Internet \cite{huberman1999internet}, Social Friendship networks \cite{wasserman1994social}, Collaboration networks \cite{newman2001structure}, etc. In these varieties of the networks, each node possesses some unique characteristics that are used to define its importance based on the given application context. In various real-life applications, we need to identify highly influential or important nodes; for example, if we want to set up a service center for the public, then which location is best for this, or if we want to provide free samples of the product then to whom we should give it, or if we want to identify popular people in society then how to find them? The importance of a node changes based on the given application context. A high degree node can be influential in spreading the information in the local neighborhood, but it will not be able to make the information viral globally if it is not connected with the other influential/core nodes \cite{kitsak2010identification}. Similarly, a high betweenness node will be important to spread the information across the communities, but it might not be influential locally based on its connections in the local community. Researchers have studied these phenomena and defined various centrality measures to identify important nodes based on the application requirements like degree centrality \cite{shaw1954some}, semi-local centrality \cite{chen2012identifying}, closeness centrality \cite{sabidussi1966centrality}, betweenness centrality \cite{freeman1977set}, eigenvector centrality \cite{stephenson1989rethinking}, katz centrality \cite{katz1953new}, PageRank \cite{brin1998anatomy}, and so on.

These centrality measures can be categorized as local centrality measures and global centrality measures. Centrality measures that can be calculated using local information of the node are called local centrality measures like degree centrality, semi-local centrality, etc. But the computation of global centrality measures, such as closeness centrality, betweenness centrality, eigenvector centrality, coreness centrality, PageRank, etc., requires the entire structure of the network. So, these centrality measures are called global centrality measures as they can not be computed without global information. These measures have high computational complexity. 

Next, we discuss the main research problems that have attracted researchers in this area.
\begin{enumerate}
\item \textbf{Extensions:} 
In 1977, Freeman proposed three main centrality measures to identify the importance of nodes based on its local and global connectivity \cite{freeman1977set}. The proposed definitions were applicable for undirected and unweighted networks. But these unweighted networks are not enough to convey the complete information of the system. The different complex systems are represented using a variety of networks like directed networks, weighted networks, multiplex networks, and so on. These networks require redefining the centrality metrics to measure the importance of nodes. We will discuss these extensions in the current report.

\item \textbf{Approximation Algorithms:} The computation of global centrality measures takes more time in large scale networks due to their high computational complexity. So, researchers have proposed approximation methods to fast compute centrality values in real-world large scale networks. 
These approximation methods can be efficiently used to compare two nodes, where we don't need actual centrality values.
\item \textbf{Update centrality values in dynamic networks:} Real-world networks are highly dynamic. It will be very costly to compute global centrality measures every time the network is updated like addition or deletion of nodes or edges. So, researchers have proposed efficient methods to update centrality values in the dynamic networks whenever there is a change in the network.
\item \textbf{Identification of top-k nodes:} In many applications, we are only interested in identifying the top few nodes like identifying top-k important nodes in the Internet system to provide instant backup, selecting top-k people to provide free samples, etc. There are methods to identify top-k nodes without computing the centrality value of all nodes. 
It reduces the overall complexity of the method.
\item \textbf{Ranking of a node:} Main objective to define centrality measures is to rank nodes. There is not much work in this direction. In one of our previous works, we have proposed a method to estimate the degree rank of a node using local information. To measure the global rank of a node using local information based on other centrality measures is still an open research question.
\item \textbf{Applications:} The applications of centrality measures are highly dependent on the requirements. The discussion on each centrality measure is concluded with its applications. We also discuss which centrality measure can be applied to what types of applications in the Applications section.
\item \textbf{Others:} A few other related works on global centrality measures are like the computation of centrality measures in distributed networks, parallel algorithms to fast compute centrality values, correlations of different centrality measures, hybrid centrality measures, and so on. The works that study the correlation of different centrality measures include \cite{valente2008correlated, tallberg2000comparing}. We will discuss these in brief in section 3.
\end{enumerate}

The brief categorization of research work on centrality measures is shown in Figure 1. Other research directions include the proposal of hybrid and application-specific centrality measures. These centrality measures work better for some specific types of networks.

\begin{figure}[]
\centering
\includegraphics[width=14cm]{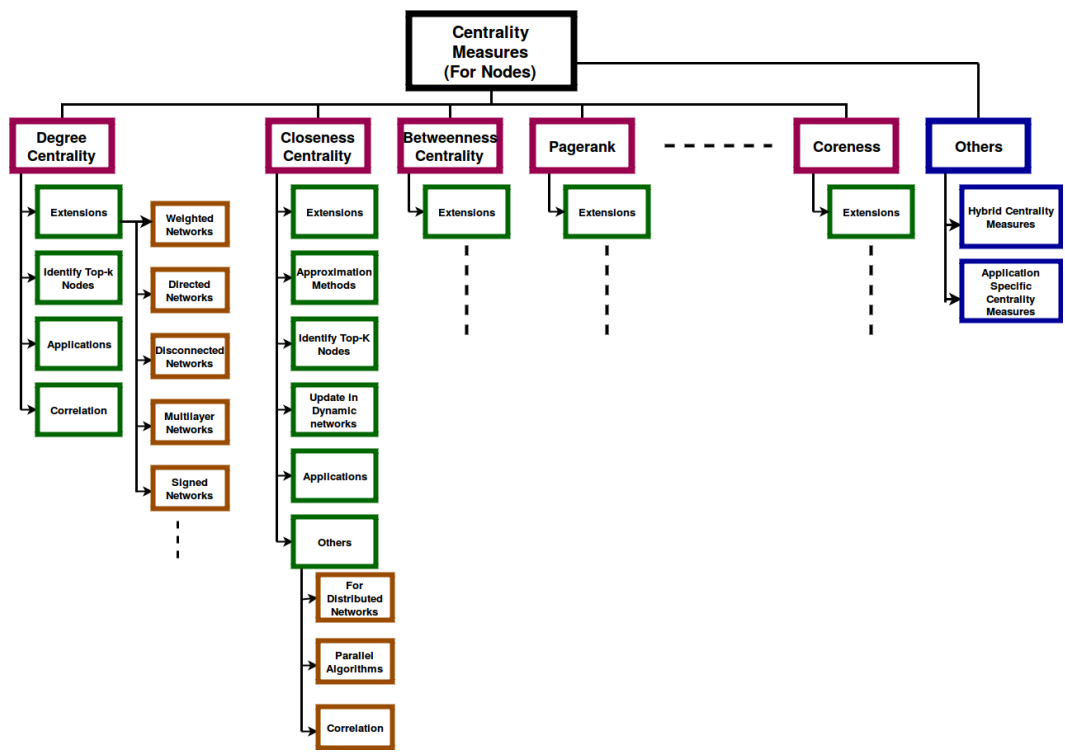}
\caption{Categorization of research work on Centrality Measures}
\end{figure}

Centrality measures also have been defined for a group of nodes to measure how central the group is with respect to the given network. For example, the closeness centrality of a group of nodes can be computed to understand how close these nodes are in the given network. Similarly, the coreness of a group of nodes can be computed to measure how tightly knit these nodes are with each other and with the rest of the network.

Like nodes, the complex networks also have unique characteristics. 
Few simple parameters to compare two networks are like their densities, diameters, the rate of changes, clustering coefficient, assortativity, and so on. Apart from these simple parameters, there are some other methods that can be used to compare other complex properties of the network like core-periphery profiling that shows how central the given network is \cite{della2013profiling}. The work on centrality measures can be categorized as shown in Figure 2. In this report, our main focus is on the centrality measures defined for nodes.

\begin{figure}[]
\centering
\includegraphics[width=12cm]{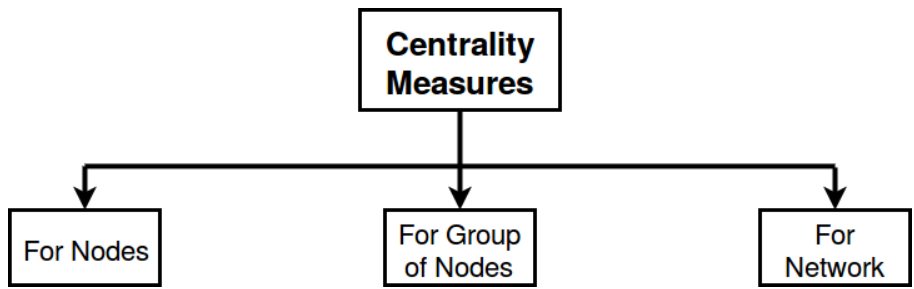}
\caption{Categories of Centrality Measures}
\end{figure}

This report is structured as follows. In Section 2, we discuss preliminaries and definitions of centrality measures. Next, we discuss the state of the art on centrality measures. It is followed by the discussion on various real-world complex networks and the centrality measures that have been applied to study those networks. The paper is concluded in Section \ref{con}.

\section{Preliminaries}
A graph is represented as $G(V, E)$, where $V$ represents a set of the nodes, and $E$ represents a set of the edges. $n$ is the total number of nodes, and $m$ is the total number of edges in the network. $u,v,w,..$ represent nodes of the network. A binary network can be represented using a matrix $A$, where $a_{ij} =1$, if $i_{th}$ and $j_{th}$ nodes are connected with each other else $a_{ij}=0$. Similarly, a weighted network can be represented using an adjacency matrix $W$, where $w_{ij}$ represents the weight of a link between $i_{th}$ and $j_{th}$ nodes. We will use the following terminologies throughout the discussion:
\begin{itemize}
\item Adjacent Nodes: The nodes that are connected via an edge are called adjacent nodes.
\item Degree: Degree of a node $u$ is denoted by $k_u$, which represents the total number of neighbors of the node.
\item In-degree: In-degree of a node $u$ is denoted by $k^{in}_u$, representing the total number of nodes with a directed link towards $u$.
\item Out-degree: Out-degree of a node $u$ is denoted by $k^{out}_u$, that represents the total number of connections that $u$ have towards other nodes in the network.
\item Neighbor Set: $\Gamma(u)$ represents set of all neighbors of node $u$.
\item Strength $(s_u)$: In weighted networks, the strength of a node denotes the sum of weights of all edges connected to that node. It is defined as, $s_{u} = \sum_{v}w_{uv}$, and it is equivalent to degree of a node in unweighted network. In-strength and out-strength can be computed similarly to in-degree and out-degree of an undirected network.
\item Walk: $u_1, u_2, .... , u_p$ is called a walk between nodes $u_1$ and $u_p$, if $\forall i$, $u_i$ is connected to $u_{i+1}$, where $i \epsilon (1, p-1)$. Here $p-1$ is the length of the walk.
\item Path: $u_1, u_2, .... , u_p$ is called a path between nodes $u_1$ and $u_p$, if $\forall i$, $u_i$ is connected to $u_{i+1}$, where $i \epsilon (1, p-1)$. In a path, a node can not be revisited. In the circular path, only the first and last node is the same. Here $p-1$ is the length of the path.
\item Distance: $d(u,v)$ represents the distance between two nodes $u$ and $v$, if there exists a path between nodes $u$ and $v$, and there exists no path that has the length smaller than this.
\item Average distance: The average distance in a given graph represents the average number of edges that need to be traversed from one node to another node in the network. It is computed as the average of the length of the shortest path for all possible pairs of nodes in the graph. Average distance is given by
$dis_{avg} = \frac{1}{n(n-1)} \sum_{\forall u,v \& u \neq v} d(u,v)$
\item Network Diameter: The maximum distance between any pair of nodes in the given graph $G(V, E)$ is the diameter of the network.
\item Sampling: When the data set is too large, the access and analysis of the data are very slow and expensive. In this case, we use a fraction of available data to make inferences about the whole dataset. This technique is referred to as sampling. In a graph, sampling can be of different types like node sampling, edge sampling etc. In node sampling, set $V'$ is called a sampled set of nodes if $ \forall u \epsilon V' \Rightarrow u \epsilon V$. Similarly, in edge sampling set $E'$ is called sampled set of edges, if $ \forall (u,v) \epsilon E' \Rightarrow (u,v) \epsilon E$. Based on the used technique, sampling methods can be categorized as follows:
\begin{enumerate}
\item Graph Traversal: In graph traversal techniques, a node can be visited only once while sampling, for example, breadth-first traversal, depth-first traversal, etc.
\item Random Walk: Random walk is a process to traverse a network. It starts from a node, and at each step, it moves to one of its neighbors uniformly at random. In random walks, a node or an edge can be sampled more than once. Based on the requirement, the random walk can also be weighted where the probability of moving at a new node is not equal for all neighbors. It can be directly or inversely proportional to the degree or can be decided using any other function.
\end{enumerate}

\item Breadth First Traversal (BFT): Breadth-First Traversal (BFT) is an algorithm for traversing a graph. It starts from a root node and explores neighbors of the root node. At each step, it traverses through all neighbors of the nodes that were explored in the last step. The time complexity of BFS in the worst case is $O(|V| + |E|)$, since every vertex and every edge will be explored. The space complexity is $O(|V|+ |E|)$ when the graph is stored as an adjacency list, and  $O(|V|^2)$ when stored as an adjacency matrix.

\item Disconnected Graph: The graph $G(V,E)$ is called disconnected graph, if there exist at least one pair of nodes in the given graph such that there is no path between them.
\item Induced Subgraph: $H(V', E')$ is the induced subgraph of $G(V,E)$ if it satisfies following conditions:
\begin{itemize}
\item $\forall u \epsilon V' \Rightarrow u \epsilon V$
\item $\forall (u,v) \epsilon E' \Rightarrow (u,v) \epsilon E$
\end{itemize}
\item Component of a Graph: Given a graph $G(V, E)$, if $H$ is a subgraph of $G$ then it is called component of graph $G$ if,
\begin{itemize}
\item $H$ is a connected graph.
\item $H$ is not contained in any connected subgraph of $G$ which has more vertices or edges than $H$.
\end{itemize}

\end{itemize}

\subsection{Definitions}

In this section, we are going to discuss the definition of all basic centrality measures.

\begin{definition}{\textbf{Degree Centrality:}}
Degree Centrality of a node $u$ is defined as,
\begin{center}
$C_D(u)= \frac{k_u}{n-1}$
\end{center}
\end{definition}

\begin{definition}{\textbf{Closeness Centrality:}}
Closeness centrality represents the closeness of a given node with every other node of the network. In precise terms, it is inverse of the farness which in turn is the sum of distances with all other nodes. The closeness centrality \cite{freeman1978centrality} of a node $u$ is defined as,
\begin{center}
$C_C(u)= \frac{n-1}{\sum_{\forall v, v \neq u}d(u,v)}$
\end{center}
\end{definition}
In disconnected graphs, the distance between all pairs of nodes is not defined, so this definition of closeness centrality can not be applied for disconnected graphs.

\begin{definition}{\textbf{Betweenness Centrality:}}
Betweenness centrality of a given node $u$ is based on the number of shortest paths passing through the node \cite{freeman1978centrality}. This measure basically quantifies the number of times a node acts as a bridge along the shortest path between a pair of nodes. The betweenness centrality of a node $u$ is defined as,
\begin{center}
$C_B(u)= \frac{\sum_{s \neq u \neq t}\frac{\partial_{st}(u)}{\partial_{st}}}{(n-1)(n-2)/2}$
\end{center}
where $\partial_{st}(u)$ represents the number of shortest paths between nodes $s$ and $t$ with node $u$ acting as an intermediate node in the shortest path.

\end{definition}

\begin{definition}{\textbf{Eigenvector Centrality:}}
Eigenvector Centrality is used to measure the influence of a node in the network \cite{stephenson1989rethinking}. It assigns a relative index value to all nodes in the network based on the concept that connections with high indexed nodes contribute more to the score of the node than the connections with low indexed nodes.

The Eigenvector centrality for a graph $G(V,E)$ is given as,
\begin{center}
$C_E(u)= (1/\lambda) \sum A_{uv} C_E(v)$
\end{center}

where $v$ is the neighbour of $u$ and  $\lambda$ is a constant. With simple rearrangement we can express it as an eigenvector equation,
\begin{center}
$AX = \lambda X$
\end{center}
\end{definition}


\begin{definition}{\textbf{Katz Centrality:}}
Katz centrality was introduced by Katz in 1953 to measure the influence of a node \cite{katz1953new}. It assigns different weights to shortest paths according to their lengths, as the shorter paths are more important for information flow than the longer paths. Contribution of a path of length $P$ is directly proportional to $s^P$ and $s \in (0,1)$. It is defined as,
\begin{center}
$K= sA + s^2A^2 + s^3A^3 + ... + s^PA^P + .... = (I-sA)^{-1} -I$
\end{center}

where $I$ is a unit matrix, $A$ is the adjacency matrix of the graph.
\end{definition}

\begin{definition}{\textbf{Coreness:}}
A node $u$ has coreness $C_S(u) = i$, if it belongs to a maximal connected subgraph $H(V_1, E_1)$, where $\forall u, u \epsilon V_1$, and $k_u \geq i$. $k_u$ is degree of node $u$ in induced subgraph $H$.
\end{definition}

This definition of coreness is based on the K-shell decomposition method that was proposed by Seidman \cite{seidman1983network}.

\section*{Centrality Measures}
In the following subsections, we will discuss the state of the art literature of different centrality measures. We cover their variations, extensions for different types of networks, such as weighted networks, directed networks, and multilayer networks, algorithms to update centrality measures in dynamic networks, approximation algorithms, methods to identify top-k nodes, and their applications.

\section{Degree Centrality}

\begin{figure}
\begin{center}
 \includegraphics[width=10cm]{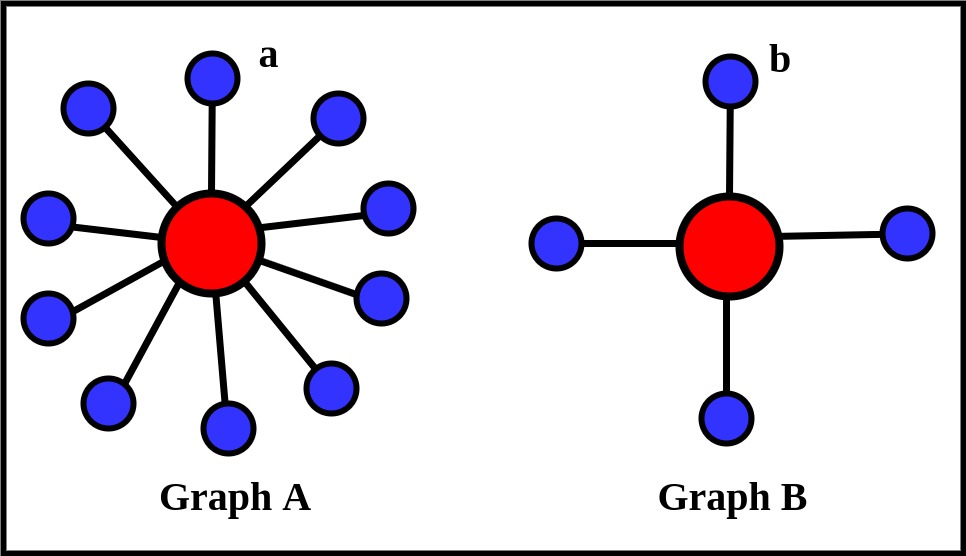}
\end{center}
\caption{Graphs A and B have two nodes a and b respectively, both having the same degree but different degree centrality.}
\label{degfig}
\end{figure}

Degree centrality term was first coined in graph theory, and it is also called degree or valence in graph theory. Degree simply denotes the number of neighbors of the node. Degree centrality is the most basic centrality measure defined in network science and it can be computed as $C_D(u)= {k_u}/{(n-1)}$. As per the definition, degree centrality is normalized using the total number of nodes. So, the first question that comes to our mind is, why is it simply not equal to the degree. Let us take an example: there are two graphs $A$ and $B$ in Figure \ref{degfig}, and in both graphs, nodes $a$ and $b$ have degree $1$. But if we analyze it, then we can simply say that node $b$ is more important, as it is connected with $25\%$ of nodes in the given network, but node $a$ is only connected with $10\%$ of the nodes. So, degree centrality is normalized using total possible connections to analyze it in a better way. It will help us to compare two nodes that belong to two different networks regardless of network size. In 1982 Grofman proposed a game-theoretic approach to measure degree centrality in social networks \cite{grofman1982game}.

\subsection{Extensions}

In directed networks, in-degree and out-degree can be used to define in-degree centrality and out-degree centrality of a node. In-degree centrality is defined as,
\begin{center}
$C_{D_{in}}(u) = \frac{k^{in}_u}{n-1}$
\end{center}
where, $C_{D_{in}}(u)$ represents in-degree centrality of node $u$, and $k^{in}_u$ represents in-degree of node $u$.

Similarly, out-degree centrality can be defined as,
\begin{center}
$C_{D_{out}}(u) = \frac{k_{out}(u)}{n-1}$
\end{center}
where, $C_{D_{out}}(u)$ represents out-degree centrality of node $u$, and $k^{out}(u)$ represents out-degree of node $u$.

In disconnected networks, the degree centrality of a node can be computed just by taking its degree in the largest connected component to that the node belongs. For isolated nodes (having degree zero), the degree centrality is zero.

Degree centrality also has been extended to weighted networks, where the strength of the node is used to define it. In weighted networks, the strength of a node $s_u$ denotes the sum of weights of all the edges connected to that node. It is defined as,

\begin{center}
$s_{u} = \sum_{v}w_{uv}$
\end{center}

In weighted networks, the degree centrality considers two important parameters: 1. strength of the node, and 2. degree of the node (total number of connections that a node has). These two parameters show two different aspects of the network. For example, in a network, if two nodes have same strength but a different number of connections, then these two nodes can not be rated equally. It is based on the requirement that more connections are preferred or fewer connections. To incorporate these both parameters, Opsahl proposed generalized degree centrality for weighted networks \cite{opsahl2010node}. It is defined as,

\begin{center}
$C_D^{w \alpha}(u)= k_u^{(1-\alpha)} \times s_u^{\alpha}$
\end{center}

where $\alpha$ is a tuning parameter that can be decided based on the requirement. If this parameter is between 0 and 1, then more importance is given to degree, whereas if it is set above 1, then more preference is given to the strength of the node. But still, it is difficult to determine the exact value of $\alpha$. Wei et al. proposed a method to select the optimal value of tuning parameter $\alpha$ \cite{wei2012degree}. Yustiawan et al. used this centrality metric to identify influential nodes in online social networks \cite{yustiawan2015degree}. This metric can be further extended to directed weighted networks, where in-degree and out-degree of the node can be considered to define in-degree and out-degree centrality, respectively.

Kretschmer used the information of weighted ties to extend degree centrality \cite{kretschmer2007new}. They verify the proposed method on collaboration and citation networks. Rachman et al. created a weighted network of Twitter, where the weight of the ties is based on following/follower relationship, and the number of tweet interactions such as mention, reply, and retweet \cite{rachman2013analysis}. They show that the Kretschmer method can be efficiently used on Twitter network to identify influential nodes.


Degree centrality has been used for decades to identify eimportant and influential nodes in the network. By considering the simplicity of degree centrality, many extensions have been provided to better rank nodes using the degree and some other local parameters. Chen et al. studied the correlation of the clustering coefficient and the influential power of a node. They showed that the local clustering has negative impacts on the information spreading and making new connections. They analyzed the impact of clustering and proposed a local ranking algorithm named ClusterRank that considers the number of neighbors, neighbors’ influences, and their clustering coefficient \cite{chen2013identifying}. They simulated the experiment using susceptible-infected-recovered (SIR) spreading model \cite{volz2007susceptible} and showed that the clusterRank is much more efficient than other benchmark algorithms such as PageRank and LeaderRank. They also verified the proposed method on undirected networks and showed that it is more efficient than degree centrality and coreness. It runs 15 times faster than pagerank algorithm, so this can be used in real-life applications.

Yang et al. further studied the impact of triangular links on the influential power of a node and proposed a node prominence profile method that considers preferential attachment and triadic closure to determine the importance of a node \cite{yang2014predicting}. The degree centrality of a node can convey more information if the degree centrality of its neighbors is also considered. Ai et al. introduced the neighbor vector centrality metric that also considers the degree distribution of its neighbors \cite{ai2013neighbor}. Their results show that it is highly efficient to measure the importance of a node, and it can be easily computed on large networks.

In real-life applications, sometimes, we are interested in selecting a group of nodes that is highly influential. This group can be the collection of nodes that might not be highly influential individually, but as a group, they have the highest influential power. Elbirt studied the degree centrality property of a network at the node level as well as at the group level \cite{elbirt2007nature}. They further compare degree centrality across different topologies like ring lattice, small world, random network, core-periphery, scale-free, and so on.

Csato proposed generalized degree centrality based on the idea that the connections with more interconnected nodes contribute to centrality more than the connections with less central ones \cite{csato2015measuring}. Generalized degree centrality $k'_u$ for a node $u$ is defined as,
\begin{center}
$k_u=k'_u + \varepsilon \sum_{v \epsilon V \setminus u} a_{uv}[k'_u - k'_v]$
\end{center}

where $\varepsilon$ is a constant, and it is used to decide the importance of the neighbors. Generalized degree centrality is better to use than the degree centrality as it considers both the degree and role of the neighbors to measure a node's importance. It gives results similar to eigenvector centrality that we will discuss later.

From the above discussion, it is clear that degree centrality alone is not enough to give a clear picture of the importance of a node in real-world networks. In 2013 Abbasi et al. proposed hybrid centrality measures called Degree-Degree, Degree-Closeness, Degree-Betweenness by combining existing centrality measures \cite{abbasi2013hybrid}. The Degree-Degree centrality of a node is calculated by summing up the degree of its neighbors. Similarly, Degree-Closeness and Degree-Betweenness centrality of a node is calculated just by summing up the closeness and betweenness centrality of its neighbors, respectively. They also studied the correlation and importance of the proposed centrality measures with respect to other existing centrality measures. The proposed method was verified on the co-authorship network and gave a good correlation with the ranking of scholars. They have also extended the proposed method for weighted networks.

In traditional degree centrality, we only consider the number of neighbors, but we do not consider the duration of the link. This information can give us more insights to better understand the position of a node in the given network. Uddien et al. proposed a time-variant approach to degree centrality measure, called TSDC (time scale degree centrality) \cite{uddin2011time}. TSDC considers both the presence and duration of the links between nodes. They simulate the proposed approach on the doctor-patient network, where a link is established when a patient is hospitalized, and the duration of the link represents the length of the stay. If the length of the stay is smaller, TSDC can give a better explanation than traditional degree centrality. This method can be applied to real-world networks that are highly dynamic and where the strength of a link has more impact on the node.

Degree centrality is also extended to other types of networks, such as multiplex networks \cite{kivela2014multilayer}, hypergraphs \cite{berge1973graphs}, and so on. Kapoor et al. extended degree centrality for hypergraphs and proposed two centrality measures called: 1. strong tie degree centrality, and 2. weak tie degree centrality \cite{kapoor2013weighted}. They verified their methods on two real-world networks, DBLP (computer science collaboration network) and CR3 (group network of a popular Chinese multi-player online game), and showed that the proposed methods outperform degree centrality. Brodka et al. proposed multilayered degree centrality called CLDC (cross-layer degree centrality) for multilayer social networks \cite{brodka2011degree}. They check the efficacy of their method on online real-world social networks containing ten layers. The proposed method is also extended to directed networks where the cross-layer indegree centrality $(CLDC_{In})$, and cross-layer out-degree centrality $(CLDC_{Out})$ can be defined using in-degree and out-degree centrality, respectively.

\subsection{Identify Top-k Nodes}
In recent years, the attention of researchers has been shifted towards the more challenging task of identifying top-k nodes in a given network. It mainly focuses on finding top-k nodes with high accuracy and without having knowledge of the entire network. The main challenge is that the complexity of the proposed methods should be many folds lesser than the actual method so that they can be applied in real-life applications. Lim et al. proposed a sampling technique to estimate top-k central nodes in the network \cite{lim2011online}. To verify the accuracy of the proposed method, the authors investigated two types of errors: 1. sampling error and 2. identification error. When a node is not sampled in the top-k most central nodes identified by the sampling algorithm, then this is called the sampling error. Similarly, if a sampled top-k node is not identified as a top-k influential node, it is referred to as the identification error. They showed that among the analyzed methods, random walk yields low sampling error. The proposed technique can also be used for other centrality measures. They further showed that the degree centrality could be used in the sampling process to reduce identification error for closeness and betweenness centrality.

\subsection{Ranking}

The main objective to define centrality measures is to rank nodes in the given network \cite{saxena2017global}. The degree centrality rank of a node $u$ can be computed as, $R(u) = \sum_{v} X_{uv} + 1$, where $X_{uv} = 1,$ if $C_D(v) > C_D(u)$, otherwise $X_{uv} = 0$. A node having the highest centrality value is ranked $1$, and all nodes having the same centrality value will have the same rank \cite{saxena2015estimating, saxena2015rank}. Currently, we follow a two-step process to compute the rank, 1. compute centrality value of all the nodes, and 2. compare them to measure the rank of the node. This method requires the entire network to measure the rank of a single node. The size of real-world networks is increasing very fast, and they are highly dynamic. This motivates researchers to propose efficient methods to estimate the centrality rank of a node without computing the centrality value of all the nodes. 

Saxena et al. \cite{saxena2016estimating} proposed a method to estimate the degree rank in real-world scale-free networks, and proved that the expected degree rank of a node $u$ can be computed as, $E[R_{G}(u)] \approx n \left( \frac{k_{max}^{1-\gamma} - (k_{u}+1)^{1-\gamma}}{ k_{max}^{1-\gamma} - k_{min}^{1-\gamma} } \right)  + 1$, where $n$ denotes the network size, $k_{max}$ and $k_{min}$ denote the maximum and minimum degree in the network respectively, and $k_u$ denotes the degree of node $u$. The authors also proposed methods \cite{saxena2017observe, saxena2017degree} to estimate degree ranking using different sampling techniques, such as uniform sampling, random walk \cite{lovasz1993random}, and metropolis-hastings random walk \cite{metropolis1953equation}. The sampling methods were used to collect a small sample set of the nodes for the rank estimation, and the results showed that $1\%$ samples are adequate to estimate the degree rank of a node with high accuracy. The proposed methods were simulated on both synthetic as well as real-world scale-free networks, and the accuracy was evaluated using absolute and weighted error functions. The proposed methods were further extended to estimate the degree rank in random networks \cite{saxena2018estimating}.

\subsection{Applications}

Imamverdiyev et al. used degree centrality to understand the dynamic relationships among a group of female BSc. students \cite{imamverdiyev2010longitudinal}. They analyzed how do these relationships change when there are some special events and how they stabilized again. It is also important in a network to understand how the importance of different nodes changes with time. Holme et al. studied the dynamics of networking agents using degree and closeness centrality measures \cite{holme2006dynamics}. Agents use local information based strategies to improve their chance of success, that is directly related to their position in the network. The success of an agent increases with its closeness centrality and size of the connected component it belongs to, while it decreases with its degree. Thus the score function is defined as,

\begin{center}
$s(u)=\left\{\begin{matrix}
C_C(u)/k_u & ,if &  k_u > 0 \\
0 & ,if & k_u = 0
\end{matrix}\right.$
\end{center}

They also showed that the network is stabilized itself between a fragmented state and a state having a giant component, and the level of fragmentation is decreased as the size of the network is increased. 

In online social networks, the spread of a meme depends on the influential power of the source node. Various parameters have been considered to identify these influential nodes. Different centrality measures are highly dependent on the network parameters and its structure. Maharani et al. observed the difference between the top ten influential nodes using degree centrality and eigenvector centrality on Twitter \cite{maharani2014degree}, and the result showed that there is a significant difference among top-10 influential users. Wambeke et al. studied the underlying social network of trades using degree and eigenvector centrality \cite{wambeke2011using}. They analyze the social network data of a project over seven months and show that such analysis can be very helpful for the management team to understand the underlying activities going on the network. Martino et al. used degree and eigenvector centrality to study attention-deficit/hyperactivity disorder (ADHD) on the brain connectivity network \cite{di2013shared}. Degree centrality has also been used to identify influential spreaders to propagate true information in the network to minimize the negative impact of fake news. Budak et al. \cite{budak2011limiting} showed that selecting a minimal group of users having the highest degree to disseminate true information in the network gives good results in mitigating bad information. 

In this section, we have discussed various extensions and applications of degree centrality in real-world networks. These are mainly used due to their simplicity and fast computation in dynamic online networks like Facebook, Twitter, WWW, and so on. Degree centrality of a node conveys its importance in the local neighborhood, but it is not able to depict its importance with respect to various other global parameters of the network. To understand the impact of network structure on the importance of a node, researchers have defined many other centrality measures. We are going to discuss those in the following sections.

\section{Closeness Centrality}

In many real-world applications, information travels through the shortest paths. In such kinds of applications, a node will be highly influential if it has a shorter distance to other nodes. This network property is captured by closeness centrality. The closeness centrality of a node denotes how close a node is in the given network. It is inversely proportional to the farness of the node. Freeman defined the closeness centrality as, $C_C(u)= \frac{n-1}{\sum_{\forall v, v \neq u}d(u,v)}$.

One similar centrality measure is graph centrality, that was proposed by Hage and Harary \cite{hage1995eccentricity}. It is defined as,

\begin{center}
$C_G(u) = \frac{1}{max_{v \in V}d(u,v)}$
\end{center}

\subsection{Extensions}

Researchers have also extended the closeness centrality definition for different types of networks. In weighted networks, closeness centrality of a node $u$ can be defined as,

\begin{center}
$C_C^w(u) = \left [\sum_{\forall v}d^w(u,v) \right ]^{-1}$
\end{center}

where $v \epsilon V$ and $d^w(u,v)$ is the shortest weighted distance between nodes $u$ and $v$. Ruslan et al. used arithmetic mean approach and extended closeness centrality for weighted networks \cite{ruslan2015improved}. The proposed centrality metric was simulated on synthetic networks. They showed that the proposed method gives better results to identify influential nodes in weighted networks.

Du et al. proposed effective distance closeness centrality (EDCC) for directed networks that can also be applied to undirected, weighted, and unweighted networks \cite{du2015new}. The effective distance was proposed by Brockmann to analyze disease spread \cite{brockmann2013hidden}. In EDCC, the authors used Susceptible-Infected (SI) model \cite{zhou2006behaviors} to evaluate the performance of the proposed method. SI model is used to measure the influential power of a node. This model is used to simulate information flow in a given complex network. In this model, all nodes are considered uninfected in the starting. Then few nodes are infected to understand the flow of the information spread. Once a node is infected, its status is changed to infected. An infected node gets one chance to infect its neighbors in each step. When a node is infected, it can infect any of its neighbors with some fixed probability in the next iteration. Thus this model perfectly captures the exponential information flow in real-world networks. The authors used this model to rank nodes based on the information flow. The efficiency of the proposed method is verified on four real-world networks. Brandes and Fleischer proposed the idea that the information spreads like an electric current in a network as it does not spread only using the shortest paths. \cite{brandes2005centrality1}. They proposed a variation of closeness centrality based on that and also proposed a faster method to compute the same.

In a disconnected network, the basic definition of closeness centrality is not able to rank nodes properly. The traditional closeness centrality is defined as,

\begin{center}
$C_C(u) \propto \left [\sum_{v}d(u,v) \right ]^{-1}$
\end{center}
If the graph is disconnected, then at least one term in the summation will be $\infty$, so the summation will be $\infty$, and the closeness centrality of all nodes will be $0$. One solution that we can use is, compute the closeness centrality of each node with respect to the connected component to which this node belongs. But it ignores a lot of other information that is present in the network.

In 2001, Latora et al. proposed closeness centrality for disconnected networks \cite{latora2001efficient}. It is defined as,
\begin{center}
$C_C(u) = \sum_{v} {\frac{1}{d(u,v)}} $
\end{center}

In 2006, Dangalchev proposed another definition of closeness centrality that can be used for disconnected graphs \cite{dangalchev2006residual}. It is defined as,
\begin{center}
$C_C(u) = \sum_{v} {\frac{1}{2^{d(u,v)}}} $
\end{center}
This method provides an easy way to compute closeness centrality than the traditional way. Yang et al. showed that all provided extension of closeness centrality for disconnected graphs are not true extensions \cite{yang2011extensions}. They do not rank nodes in the same way as the closeness centrality does. In 2009, Rochat proposed a harmonic centrality index that is an alternative to the closeness centrality index for disconnected networks \cite{rochat2009closeness}. The author showed that the results are the same on connected networks with the same computational complexity. The main benefit is that it can be used for disconnected networks.

In real-world social networks, a node can be part of multiple communities that gives birth to overlap community structure. Tarkowski et al. defined closeness centrality for the networks having overlapped community structure \cite{tarkowski2016closeness}. They used a cooperative game-theoretic approach to compute the closeness centrality of a node in polynomial time complexity. They also verified their results on Warsaw public transportation networks. Still, the use of game-theoretic approaches for other centrality measures like betweenness centrality, eigenvector centrality, is an open problem.

Barzinpour et al. extended closeness centrality for multilayer networks \cite{barzinpour2014clustering}. He explained how inter-connections and intra-connections of layers could be used to get the true position of a node based on closeness centrality.

Yu et al. combined closeness centrality and degree centrality to identify important nodes in the given network \cite{yunode}. The algorithm is defined as,
\begin{enumerate}
\item Calculate the shortest distance matrix of the given graph.
\item Calculate node importance evaluation matrix using degree centrality. This is defined as,
\begin{center}
$E=\begin{bmatrix}
k_1 & a_{12}k_1k_2/2m & \cdots & a_{1n}k_1k_n/2m\\
a_{21}k_1k_2/2m & k_2 & \cdots & a_{2n}k_2k_n/2m \\
\vdots  & \vdots & \vdots & \vdots \\
a_{n1}k_1k_n/2m & a_{n2}k_2k_n/2m & \cdots  & k_n
\end{bmatrix}$
\end{center}

where $k_i$ is degree of node $i$.
\item Improve node importance evaluation matrix using closeness centrality.
\begin{center}
$H=\begin{bmatrix}
k_1C_1 & a_{12}k_1k_2C_2/2m & \cdots & a_{1n}k_1k_nC_n/2m\\
a_{21}k_1k_2C_1/2m & k_2C_2 & \cdots & a_{2n}k_2k_nC_n/2m \\
\vdots  & \vdots & \vdots & \vdots \\
a_{n1}k_1k_nC_1/2m & a_{n2}k_2k_nC_2/2m & \cdots  & k_nC_n
\end{bmatrix}$
\end{center}

where $C_i$ is closeness centrality of node $i$.
\item Calculate importance of each node using given formula that is defined as,
\begin{center}
$I_i = \sum_{j=1}^{n}H_{ij} = k_iC_i + \sum_{j=1,i\neq j}^{n} H_{ij}$
\end{center}
\end{enumerate}

Thus, the proposed centrality metric considers both local and global information of the node. The proposed method is verified on real-world networks like the Advanced Research Project Agency (ARPA) network and AIDS network as well as synthetic networks.

\subsection{Approximation Methods}

The complexity to compute the closeness centrality of all nodes is $n$ times the complexity of BFT. Eppstein et al. proposed a randomized approximation algorithm to fast compute closeness centrality in weighted networks \cite{eppstein2004fast}. The proposed method calculates the centrality of all vertices in $O(m)$ time, where $m$ is the number of edges. Cohen et al. proposed a method to approximate closeness centrality for directed and undirected networks \cite{dellingcomputing}. The proposed approach is a combination of the exact and sampling method. In the sampling method, few nodes are sampled uniformly at random, and BFT is executed from these nodes. The execution of these BFTs will determine the shortest paths of all nodes with the sampled nodes. In the hybrid approach, while calculating the closeness centrality of a node, a threshold distance is decided. Now to compute the closeness centrality of a node, its exact distance is computed with all nodes within the threshold distance. For all nodes which fall outside the threshold distance, two approaches are used: 1. If the node is a sampled node, its exact distance is already known; otherwise 2. approximation method is used to approximate the shortest distance of the node. Thus, the closeness centrality of a node is calculated in linear time complexity using this approach.

Rattigan used the concept of network structure index (NSI) to approximate the values of different centrality measures that need to identify the shortest paths in the given network \cite{rattigan2006using}. Shi et al. developed a software package gpu-fan (GPU-based Fast Analysis of Networks) for fast computation of centrality measures in large networks \cite{shi2011fast}. This method can be used for other centrality measures if they also use the computation of shortest paths like betweenness centrality, eccentricity centrality, and stress centrality. Eppstein and Wang proposed an approximation to compute closeness centrality \cite{eppstein2001fast}. The proposed method estimates the centrality value of all nodes in $O(n)$ time with $(1 + \epsilon)$ linear approximation factor. Some other approximation methods for closeness centrality include \cite{chan2009fast, brandes2007centrality, pfeffer2012k}.

\subsection{Update in Dynamic Networks}

Real-world networks are highly dynamic, and their structure keeps changing at every single moment by addition and removal of nodes or edges. Kas et al. proposed closeness centrality for dynamic networks \cite{kas2013incremental}. Whenever there is any addition, removal, and modification of nodes or edges, we can make the set of affected nodes and update all pair shortest paths using that. In 2013, Yen also proposed an algorithm called CENDY (Closeness centrality and avErage path leNgth in DYnamic networks) to update closeness centrality when an edge is updated \cite{yen2013efficient}. They also used this approach to propose a method to update the average path length of the network just by computing a very small number of shortest paths. Sariyuce et al. proposed a method to update closeness centrality using the level difference information of breadth-first traversal \cite{sariyuce2013incremental}. They also used biconnected component decomposition, spike-shaped shortest-distance distributions, and identical vertices techniques to improve their results. The proposed method gives improvement by a factor of 43.5 on small networks and 99.7 on large networks than the traditional non-incremental algorithm \cite{brandes2001faster}.

\subsection{Parallel and Distributed Computation}

In 2006, Bader et al. proposed a parallel algorithm to compute closeness centrality, where it executes a breadth-first traversal (BFT) from each vertex as a root \cite{bader2006parallel}. If there are $p$ processors, then the time taken to compute the closeness centrality of all the $V$ vertices would be $O( \frac{\left |V \right |}{p} (\left | V \right | + \left | E \right |))$. Some other network analysis libraries are also available to compute parallel closeness centrality \cite{ediger2012stinger, gregor2005parallel, lugowski2012scalable}. Lehmann and Kaufmann proposed a method for decentralized computation of closeness centrality and graph centrality \cite{lehmann2003decentralized}. The proposed method is computationally expensive for large-scale real-world complex networks. Wang et al. also proposed a distributed algorithm to compute closeness centrality \cite{wang2015distributed}. They showed that the proposed method estimates closeness centrality with $91\%$ accuracy in terms of ordering on random geometric, Erdos-Renyi, and Barabasi-Albert graphs.

\subsection{Identify Top-k Nodes}

In real-life applications, mostly, we are not interested in computing the closeness centrality of all nodes. All practical applications focus on identifying a few top nodes having the highest closeness centrality. Some of these algorithms to identify top-k nodes are discussed here. Ufimtsev proposed an algorithm to identify high closeness centrality nodes using group testing \cite{ufimtsev2014extremely}. Okamoto et al. proposed a method to rank k-highest closeness centrality nodes using a hybrid of approximate and exact algorithms \cite{okamoto2008ranking}. Olsen et al. presented an efficient technique to find k-most central nodes based on closeness centrality \cite{olsen2014efficient}. They used intermediate results of centrality computation to minimize the computation time. The proposed method uses $O(V+E)$ additional space, and it is 142 times faster than the conventional method, where we compute the closeness centrality of each node independently.

Bergamini et al. proposed a faster method to identify top-k nodes in undirected networks \cite{bergaminicomputing}. They used BFT information to determine the upper bound on closeness centrality and halt the process when top-k nodes having the highest closeness centrality have been identified. They have proposed two methods that can be used based on network properties. The first one can be used for small-world graphs with low diameter, and the second one can be used for the graphs having a high diameter. The proposed method outperforms the state of the art methods to compute top-k nodes.

Wehmuth et al. proposed a method named DACCER (Distributed Assessment of the Closeness CEntrality Ranking) to estimate the closeness ranking of nodes using local information \cite{wehmuth2012distributed}. They have shown that the DACCER rank is highly correlated with closeness centrality rank for both real-world and synthetic networks. The DACCER centrality is computed using the local neighborhood volume of the node. It is defined as,
\begin{center}
$Vol(H_h^u) = \sum_{v \epsilon H_h^u} k_v$
\end{center}

where $k_v$ is the degree of node $v$, and $h$ is the level of breadth-first traversal (BFT). $H_h^u$ denotes the set of all nodes that belong to $h$ level BFT of node $u$. By definition, the volume of a node gives us the sum of degrees of all nodes that belong to the h-level BFT of the given node. The results show that if we take $h=2$, then the ranking is highly correlated with closeness centrality ranking. This value can be easily computed for each node using 2-level BFT, and it is not required to have information about the entire network. Lu et al. extended this method and proposed MDACCER (Modified Distributed Assessment of the Closeness CEntrality Ranking) to compute closeness centrality in a parallel environment like General Purpose Graphics Processing Units (GPGPUs) \cite{lu2015mdaccer}.

\subsection{Ranking}

The classical method of computing the closeness centrality rank of a node, first computes the closeness centrality value of all nodes, and then compare its closeness value with others to determine the closeness rank of the node. The time complexity of the first step is $O(n \cdot m)$ to compute the closeness centrality of all nodes; for the second step, it is $O(n)$ to compare the centrality value of the given node with all other nodes. So, the overall time complexity of this process is $O(n \cdot m)+O(n)= O(n \cdot m)$, which is very high. This high complexity method is infeasible to use in real-life applications of large size networks. Saxena et al. \cite{saxena2017faster, saxena2017fast} studied the structural properties of closeness centrality in real-world networks and observed that the reverse rank\footnote{In the reverse ranking, the node having the lowest closeness value has the highest rank, namely $1$, and the node having the highest closeness value has the lowest rank.} versus closeness centrality follows a \textit{sigmoid curve}. Once the parameters of the sigmoid equation are estimated, this can be used to fast estimate the closeness rank of a node. The authors proposed the methods to estimate these parameters using network properties. They further proposed heuristic methods to estimate the closeness rank of a node in $O(m)$ time that is a huge improvement over the classical method. The accuracy of the proposed methods was measured using absolute and weighted error functions, and correlation coefficients. The results showed that the proposed methods could be used efficiently for large scale complex networks of different types \cite{saxena2019heuristic}. 

\subsection{Applications}

Closeness centrality has been applied in many important research areas. Newman used closeness centrality to better understand collaboration networks \cite{newman2001scientific}. Yan et al. also used closeness centrality to understand various parameters of collaboration networks \cite{yan2009applying}. Sporns et al. used closeness centrality to identify hubs in the brain network \cite{sporns2007identification}. Park et al. proposed a method to measure closeness centrality among workflow-actors of workflow-supported social network models \cite{park2013closeness}. Kim et al. proposed an estimation driven closeness centrality based ranking algorithm named RankCCWSSN (Rank Closeness Centrality Workflow-supported Social Network) for large-scale workflow-supported social networks \cite{kim2016estimated}. They showed that the time efficiency of the proposed method is about $50\%$ than the traditional method. This method can easily be extended to weighted workflow-supported social networks.

Zhang et al. used closeness centrality to identify the community of few nodes by using community information of other nodes \cite{zhang2012inferring}. Jarukasemratana et al. proposed a community detection method using closeness centrality \cite{jarukasemratana2014community}. Ko et al. proposed the closeness preferential attachment (CPN) model to create synthetic networks using closeness centrality \cite{ko2008rethinking}. According to closeness preferential attachment law, a new node joining the network would like to make connections with other nodes that will help it to increase its closeness with the entire network. They compare the CPN model with the BA model. In the CPN model, each node tries to decrease its distance with the remaining network, but it gives birth to a longer average distance than the BA model. Other applications of closeness centrality can be looked at \cite{Hidalgo, Zhang2008StudentsIA, marsden2002egocentric, saxena2020mitigating, nasiruzzaman2011application, saqr, Liu2018, guan2020closeness}.

In this section, we have discussed the state of the art literature on closeness centrality, and next, we will discuss betweenness centrality. 

\section{Betweenness Centrality}

In 1973, Granovetter emphasized the inequality of edges in a network and introduced the idea of weak ties \cite{granovetter1973strength, granovetter1983strength}. The edges of a network can be broadly categorized as weak ties or strong ties. Strong ties represent the relationships between people who frequently contact each other, and weak ties represent the relationships having less communication frequency. Granovetter's work was the first work of its kind that distinguished between the edges of a network in some way. It simply shows that some edges play the role of bridges for information flow more frequently than others.

Similarly, in complex networks, the uniqueness of a node can be determined by its importance in the information flow in the network. This unique characteristic is captured by the betweenness centrality of the node. Betweenness centrality is based on the flow of information through nodes, so, it is also called flow centrality. It assumes that the information always flows through the shortest paths like water and electricity. It accounts for the number of shortest paths passing through a node that explains the importance of a node with respect to the information flow. The basic definition of betweenness centrality is defined as, $C_B(u)= \frac{\sum_{s \neq u \neq t}\frac{\partial_{st}(u)}{\partial_{st}}}{(n-1)(n-2)/2}$, where $\partial_{st}(u)$ represents the number of shortest paths between nodes $s$ and $t$ with node $u$ acting as an intermediate node in the shortest path. This is the extension of the stress centrality measure that was proposed by Shimbel in 1953 \cite{shimbel1953structural}. Stress centrality is based on the total number of shortest paths passing through a node; it is defined as,

\begin{center}
$C_S(u)= \sum_{s,t \in V, s \neq t} \sigma_{st}(u) $
\end{center}

where, $\sigma_{st}(u)$ is the number of shortest paths from $s$ to $t$ that passes through $u$.

The time complexity of betweenness centrality is very high $(O(m^3))$, as it counts the total number of shortest paths passing through a node for each pair of nodes present in the network.

In 2001, Brandes proposed a fast method to compute the betweenness centrality of all nodes. The proposed algorithm takes $O(n+m)$ space, and $O(nm)$ and $O(nm + n^2 logn)$ time on unweighted and weighted networks respectively \cite{brandes2001faster}. The proposed method executes BFT from a node and counts the number of shortest paths passing through a node using this information. Thus, the total number of shortest paths passing through a node is counted by executing BFT once from each node. The complexity of BFT on an undirected network is $O(m)$, so the complexity of betweenness centrality is $O(nm)$.

\subsection{Extensions}


Betweenness centrality is also extended based on the specific application requirements. Freeman proposed a family of new centrality measures based on the betweenness centrality \cite{freeman1977set}. These measures can be used for both connected and disconnected networks. Brandes also proposed some variations of standard betweenness centrality metric called endpoints, proximal betweenness, k-betweenness, length-scaled betweenness, linearly scaled betweenness, edge betweenness, group betweenness, q-measures, stress centrality, and load centrality \cite{brandes2008variants}. He also extended betweenness centrality for valued networks and multigraphs. Brandes and Fleischer proposed a variation of betweenness centrality based on the assumption that the information spreads like an electric current in the network \cite{brandes2005centrality1}. As the computation of this centrality measure is very costly on large scale networks, so they also proposed an approximation method for the same.

\subsection{Approximation Methods}

Betweenness centrality only considers the number of shortest paths passing through a node. It assumes that the information always flows through the shortest paths. But it might not always true in real-world scenarios. Information can also pass through longer paths with some probability. Newman considered this fact and proposed a betweenness measure that considers all paths, but more importance is assigned to shorter paths \cite{newman2005measure}. The centrality of a node is computed using random walks, and it is directly proportional to how often a node is traversed while taking a random walk. He also shows that the proposed measure can be calculated using matrix inversion methods. There are several other approximation algorithms that include \cite{bergamini2015fully, geisberger2008better, riondato2014fast}.

Lehmann and Kaufmann proposed an efficient method for decentralized computation of betweenness centrality and stress centrality \cite{lehmann2003decentralized}.

\subsection{Update in Dynamic Networks}

Betweenness centrality has a high computational cost than other centrality measures. In dynamic networks, it is not feasible to recompute the centrality values if the network is updated. Some methods have been proposed to update betweenness centrality in dynamic networks that we are going to discuss next.

Lee et al. proposed a method called QUBE framework to update betweenness centrality in the case of edge insertion and deletion within the network \cite{lee2012qube}. The proposed method is based on the biconnected component decomposition of the graphs. When an edge is inserted or deleted, then the centrality values within the updated biconnected component are recomputed from scratch. If the decomposition is affected due to edge insertion/deletion, the modified graph is first decomposed into biconnected components. The performance of the QUBE highly depends on the distribution of vertices to the biconnected components. In real-world networks, component size is large, so it does not reduce update time significantly. The authors show the performance of the proposed method on smaller graphs having a low density. This method performs significantly well on small graphs with a tree-like structure having many small biconnected components.

Green et al. also proposed a method to update betweenness centrality values rather than recomputing them from scratch upon edge insertions or edge deletions \cite{green2012fast}. This approach is based on storing the whole data structure used by the previous betweenness centrality update kernel. This storage helps in reducing computation time significantly as some of the centrality values will remain the same. This method uses quadratic storage space, so it is impractical to use for large size networks.

In 2015, Chernoskutov et al. proposed a method to approximate betweenness centrality values in dynamic networks \cite{chernoskutov2015heuristic}. This method contains two steps: 1. condensed the initial graph to get a smaller version, and 2. approximate betweenness centrality for smaller graph and extrapolates it to the large graph. The proposed method gives a speedup of $60\%$ for real-world networks.

\subsection{Identify Top-k Nodes}

Agryzkov et al. proposed the random walk betweenness centrality index to rank nodes based on the concept of pagerank \cite{agryzkov2014new}. First, the adapted pagerank algorithm is executed to rank nodes based on their importance. Then the final ranking of the nodes is computed using betweenness centrality ranking. They analyzed the proposed method on the real urban street network and compared it with other centrality measures. Kourtellis et al. proposed k-path centrality to identify nodes having high betweenness centrality \cite{kourtellis2013identifying}. They used a randomized algorithm to identify nodes with high k-path centrality and showed that nodes with high k-path centrality also have high betweenness centrality. The proposed method executes faster than existing methods on the real world and synthetic networks. The fast estimation of betweenness centrality rank is still an open research question.


\subsection{Applications}

Newman used betweenness centrality to study collaboration networks and showed the effect of funneling \cite{newman2001scientific}. It shows that most of the shortest paths of a node pass through only the top few collaborators and remaining collaborators participate in a very small number of shortest paths. Leydesdorff used betweenness centrality to study the citation network of Journals \cite{leydesdorff2007betweenness}. Their results help us understand that betweenness centrality can measure the interdisciplinarity of journals using local citation environments. Abbasi et al. showed that the betweenness centrality is a good measure for preferential attachment than the degree and closeness centrality in collaboration networks \cite{abbasi2012betweenness}. Other applications that have used betweenness centrality include \cite{sporns2007identification, marsden2002egocentric, yan2009applying, Dawson, Nurmela, kimura2009blocking}.

Brandes et al. studied the dependency of closeness and betweenness centrality \cite{brandes2016maintaining}. Before this work, researchers used to consider both of these centrality measures independently. This was the first work of its kind that show the inter-dependency of both centrality measures mathematically. Real-world scale-free networks can be categorized into three categories based on the average nearest neighbor degree: 1. assortative networks, 2. disassortative networks, and 3. neutral networks \cite{newman2002assortative}. In assortative networks, a node with a high degree tends to be connected with other nodes having high degrees. Few examples of assortative networks are social networks, co-authorship networks, actor networks, and so on. In disassortative networks, a node with a high degree tends to be connected with other nodes having low degrees and vice versa, for example, Internet network, WWW network, biological networks, and so on. Goh et al. studied the correlation of betweenness centrality in scale-free networks \cite{goh2003betweenness}. Results show that the betweenness centrality correlation behaves the same in disassortative and neutral networks. But in assortative networks, it shows a different pattern. In assortative networks, a node is connected to other nodes having the same influential power. These results are highly important in understanding information dynamics in different networks.


In this section, we have discussed betweenness centrality, its fast computation algorithms, approximation methods, and its extensions. There has not been any significant work to identify top-k nodes in betweenness centrality as the complexity to compute one node's centrality is equivalent to compute the centrality of all nodes. Betweenness centrality is highly applicable in real-life applications, and we have discussed it in the later part of the section.

\section{PageRank Centrality}


PageRank is the key parameter to measure the success of search engines, which helps to find out top results for the given search query. Pagerank is a global centrality measure that needs the entire network to measure the importance of one node. It measures the importance of one node based on the importance of its neighbors. Thus it is an iterative process that uses global information to estimate where you stand in the network. The first method to compute PageRank was proposed by Brin and Page in 1998 while developing the ranking module for the prototype of Google \cite{brin1998anatomy}. Various other methods also have been proposed to compute the pagerank value of a node quickly.

In dynamic big real-world networks, we use the random walk based method to estimate the rank of a node. To compute the pagerank, a few crawlers are started to walk on the network. Initially, the counter for all nodes is set to zero. When a crawler reaches a node, it increases its counter by one and moves to one of its neighbors uniformly at random. While taking random walks, there can be situations when a crawler can be stuck in some part of the network or in a community or be stuck on a node with no outgoing links. To handle such conditions, the teleportation facility is used. In teleportation, when a crawler reaches a node, then with probability $q$ (where $0 < q < 1$), it selects a node uniformly at random in the entire network and jumps to it, and with probability $(1-q)$ it moves to one of its neighbors randomly. It is observed that $q \approx 0.15$ gives good results in real-world networks.

Pagerank of a node is defined as,

\begin{center}
$P(u)=\frac{q}{n}+(1-q)\sum_{v:v\rightarrow u} {P(v)/k^{out}_v}$
\end{center}

where $n$ is the total number of nodes in the network, $q$ is teleportation factor, and $k^{out}_v$ is the out-degree of node $v$. $v \rightarrow u$ shows a link from $v$ to $u$. Thus the pagerank value $P(u)$ shows the probability to find the crawler at node $u$ when the complete process converges to a stable state.



\subsection{Extensions}

Pagerank has also been extended for different types of networks. Xing and Ghorbani \cite{xing2004weighted} extended the page rank for weighted networks and proposed the weighted PageRank algorithm (WPR). The WPR method considers the importance of the links and distributes rank scores based on the popularity of the nodes. The results show that the WPR method performs better than the classic PageRank method in terms of finding the relevant pages to a given search query. Pagerank has also been extended for temporal networks \cite{lv2019pagerank, rozenshtein2016temporal, hu2015temporal}, multilayer networks \cite{cheriyan2020improved, tu2018novel, pedroche2016biplex, lv2019nodes}, and hypergraphs \cite{tran2019pagerank}.

Fiala \cite{fiala2012time} extended the pagerank for bibliographic networks as in these networks, we have temporal and meta-information about the citations and authors. The proposed methods weigh citations between authors based on the information from the co-authorship network, and the methods were tested on the Web of Science dataset of computer science journal articles to determine the most prominent computer scientists in the period of 1996–2005.

\subsubsection*{Customized Pagerank}

Many research papers talk about the customized ranking and how to improve users' experience on the search engines. These specialized rankings are suitable for many particular applications. The main underlying idea of specialized ranking is based on the concept that the page importance is not absolute, but it depends on the particular needs of a search engine or a user. For example, If a user is searching for the list of all top institutes, then the home page of all these institutes will not be so important. He would like to get a page that has consolidated information of all top institutes with their rankings. Specialized pagerank also can consider the user history and preferences to decide the order to display search query results. For example, different institutes might be interested in customizing the ranking algorithm for the specific environment. Scarselli et al. proposed a neural network model to compute customized page ranks in World Wide Web \cite{scarselli2005graph}. Many approaches have been proposed for specialized page ranking based on the topic, user, or search query \cite{diligenti2004unified, haveliwala2002topic, jeh2003scaling, richardson2001intelligent, tsoi2003adaptive}. Honglun et al. have compared various techniques to get personalized rankings and have written a survey on the same \cite{honglun2013efficient}.

\subsection{Approximation Methods}

Amento et al. analyzed the correlation between pagerank and in-degree based on five queries \cite{amento2000does}. They show $60\%$ average precision as observed by the human subjects. There are also some other studies that show the correlation of pagerank with in-degree in web network \cite{pandurangan2002using, donato2004large, nakamura2003large}. The plot between pagerank and in-degree follows power-law distribution with a broad tail having power-law exponent $\gamma \approx 2.1$. Grolmusz showed that the pagerank in an undirected graph is not directly proportional to the degree \cite{grolmusz2015note}. They proposed an upper and a lower bound for the pagerank distribution and explained necessary and sufficient conditions for the PageRank to be proportional to the degree.

Litvak et al. performed experiments and observed that the pagerank and in-degree both obey the power law with the same exponent \cite{litvak2006degree}. They presented a mathematical model using a stochastic equation to explain this phenomenon. They also showed that the tail behavior of the PageRank and the in-degree differs only by a multiplicative factor, and derived a closed-form expression for the same. They have further worked to propose Monte Carlo methods to identify top-k personalized PageRank lists \cite{avrachenkov2010monte}. There are few other works that propose efficient approaches to identify top-k nodes based on the requirement \cite{cao2010retrieving}.

In 2008, Fortunato et al. used the mean-field theory to understand the correlation of pagerank with in-degree of the node \cite{fortunato2006approximating}. They showed that the pagerank is directly proportional to the in-degree, modulo an additive constant. They also showed that the global ranking $R(P)$ of a node based on pagerank $P$ could be defined as,
\begin{center}
$R(P) \approx AP^{-\beta}$
\end{center}

where $\beta = \gamma - 1 \approx 1.1$. $\gamma$ is the power law exponent of pagerank distribution and $A$ is a proportionality constant.

This complete process includes following steps:
\begin{enumerate}
\item Compute pagerank of a node using following equation:
\begin{center}
$P(k) = \frac{q}{n} + \frac{1-q}{n}\frac{k_{in}}{\left\langle k_{in}\right\rangle }$
\end{center}
This gives the average PageRank of all nodes having degree $k$.
\item Compute the global rank $R(P)$ of the node
\item A page with global ranking $R$ can be placed at any position in the hit list of length $h$ based on the query. We can compute local ranking $r$ of the node as,
\begin{center}
$r=R\frac{h}{n}$
\end{center}
\end{enumerate}
Using this approach, we can compute pagerank of a node if we know its in-degree. The combined expression to calculate local rank of a node can be written as,
\begin{center}
$r=\frac{Ah}{(\frac{q}{n} + \frac{1-q}{n}\frac{k_{in}}{\left\langle k_{in}\right\rangle })^{1.1} n}$
\end{center}

Broder et al. proposed a framework to compute random walk based pagerank values for web networks \cite{broder2006efficient}. The proposed method shows the speedup of 2.1, and the spearman correlation of ranking with pagerank is 0.95. Kamwar et al. proposed a technique to rank nodes in large real-world directed networks \cite{kamvar2003exploiting, kamvar2007methods}. They partitioned the link matrix into blocks, and the local ranking of each node is calculated in the corresponding block. Then the block level rank is used to estimate the global rank of the node. This method will give fast results as we can use a distributed computing environment to calculate local ranks in different blocks. Shariaty also analyzed the impact of neighbors on the pagerank of a node and proposed an approximation method based on local neighborhood information \cite{shariaty2011local}. There are some other works that use local information to estimate pagerank value \cite{chen2004local}.

Richardson et al. proposed a machine learning based approach to approximate static pagerank values using user history and other static features \cite{richardson2006beyond}. Liu et al. used two sampling methods, 1. direct sampling and 2. adaptive sampling, to approximate google pagerank values \cite{liu2015fast}. The direct sampling method samples the transition matrix once and uses the sample directly in PageRank computation, whereas the adaptive sampling method samples the transition matrix multiple times. In adaptive sampling, the sample rate can be adjusted iteratively as the computing procedure proceeds. They have simulated the methods on six real-world datasets, and results show that the proposed methods can achieve higher computational efficiency. Luh used a latent semantic analysis approach to approximate Google pagerank \cite{luh2011approximating}.

In pagerank algorithm, we can also modify the sampling technique to converge the values faster or to get the application specific results. Boldi et al. analyzed the crawl strategies and studied whether the results obtained by partial crawling can be used to represent global ranking or not \cite{boldi2004your}. They analyze when the crawling process can be stopped to announce the result that is very close to the actual ranking of the nodes. They performed the experiment on real-world networks and compared the ranking using Kendall's coefficient \cite{kendall1948rank}. Results show that if the sampling strategy computes the pagerank quickly, then it is badly correlated with the actual ranking. But the results are opposite for synthetic random graphs.

Keong et al. proposed a modified random surfer model, which makes the number of iterations required to compute PageRank more predictable \cite{keong2011pagerank}. They showed that 30 iterations are enough to accumulate the total PageRank up to 0.992, and 50 iterations are enough to accumulate the total PageRank up to 0.9997 theoretically. Borgatti et al. studied the effect of sampling on different centrality measures like degree centrality, closeness centrality, betweenness centrality, and eigenvector centrality \cite{borgatti2006robustness}. They showed that the accuracy of centrality measures decreases as the sample size decreases. Maiya et al. also studied the impact of sampling techniques to identify highly influential nodes \cite{maiya2010online}.

Haveliwala presented convergence results for deciding the number of iterations that are required to get stable pagerank values \cite{haveliwala1999efficient}. Yu et al. proposed IRWR (Incremental Random Walk with Restart) approach to update the pagerank in $O(1)$ time in dynamic networks \cite{yu2013irwr}. Mainly they proposed a fast incremental algorithm that shows high efficiency and exactness for computing proximities whenever an edge is updated. Sarma et al. proposed random walk based distributed algorithms for computing pagerank in directed and undirected graphs \cite{sarma2015fast}. The first approach takes $O(logn/q)$ rounds in both directed and undirected networks, where $n$ is the total number of nodes and $q$ is the teleportation factor. They also proposed a faster algorithm for undirected networks that takes $O(\sqrt{logn}/q)$ rounds. Berkhin has written a survey on pagerank computing that can be referred to for further information \cite{berkhin2005survey}.

\subsection{Update in Dynamic Networks}

Like other centrality measures, various approaches have been proposed to update pagerank in dynamic networks. Pagerank is mainly used in the WWW network that is highly dynamic. Desikan et al. proposed a method to update pagerank in evolving networks based on the first-order Markov Model \cite{desikan2005incremental}. In the WWW network, whenever any change occurs, it mainly affects a small part of the graph, and the remaining large part is unchanged. The pagerank of a node is dependent on the nodes that have directed link towards it and is independent of out-going links of the node. They carefully analyzed the changed and unchanged part and their dependencies to compute the pagerank incrementally. They divided the network into two parts: 1. First partition $Q$ is such that there are no incoming links from a partition, and 2. the second part $P$ includes remaining nodes. Now, we can compute the pagerank of partition $Q$ separately and then scaled and merged it with the rest of the network to get the actual PageRank values of nodes in $Q$. The scaling is done by considering the number of nodes in both partitions. Berberich et al. proposed a normalized PageRank metric to compare two nodes, and the proposed score is robust to non-local changes in the graph, unlike the standard PageRank method \cite{berberich2007comparing}.

\subsection{Identify Top-k Nodes}

In real-world networks, the total number of nodes is very large. So, most of the time users are interested in finding top-k pages (where $k$ can be typically from 10 to 100) based on the search query and user preference. The exact ranking of lower-ranked nodes is not much important. Many researchers have looked into it and have proposed different methods to find top-k nodes \cite{fujiwara2013fast, fujiwara2012fast, zhang2015fast, avrachenkov2011quick, zhang2011preference, yu2014reverse, ilyas2008survey}. 

\subsection{Applications}

Apart from WWW network, pagerank is also used to rank nodes in different types of networks, such as citation networks \cite{ding2009pagerank, ding2011topic}, collaboration networks \cite{yan2011discovering, franceschet2011pagerank}, social networks \cite{wang2013discover}, protein interaction networks \cite{peng2014identification}, brain networks \cite{yang2020unified}, semantic networks \cite{mihalcea2004pagerank}, and so on. Some online social networks, such as Linkedin, Researchgate, ask for the users' endorsements for their special skills. A directed graph can then be created using this endorsement information, where nodes are the users, and edges represent the score of endorsements. Roses et al. used a pagerank method to rank these nodes and verified their results on a synthetic network with 1493 nodes \cite{perez2016endorsement}. Cheng et al. used pagerank and HITS to rank nodes in Journal citation networks \cite{cheng2009pagerank}. In Journal citation networks, nodes represent different journals, and there is a directed weighted edge between two nodes $(u,v)$ if journal $u$ cites journal $v$, and edge weight depends on the number of citations. In WWW network, there are no self-loops, but in the journal citation network, all journals have self-citations, and it makes self-loops in the final network. Their results present that pagerank and HITS can be used to rank journals, and it gives good ranking than the ISI impact factor. 

In this section, we have discussed pagerank, its extensions, its variations, and approximation methods to compute it. Pagerank is highly used to rank nodes when a node's importance is dependent on its neighbors. In the next section, we will discuss the coreness of the nodes representing how well a node is connected to other important nodes and also with periphery nodes in the network.

\section{Coreness Centrality}

Real-world networks have a self-regulatory evolving phenomenon that gives rise to a core-periphery structure. The hierarchical organization of the network gives birth to the core-periphery structure that coexists with the community structure. The concept of the core-periphery structure was first proposed by Borgatti and Everettee \cite{borgatti2000models}. The core is a very dense nucleus of the network that is highly connected with periphery nodes. In social networks, core nodes are the group of highly elite people of the society. Similarly, in a co-authorship network, core nodes are the pioneer researchers of the area.

Seidman \cite{seidman1983network} proposed the k-shell decomposition method to identify core nodes in an unweighted network. The k-shell algorithm is a well-known method in social network analysis to find the tightly knit group of influential core nodes in the given network. This method divides the entire network into shells and assigns a shell index to each node. The innermost shell has the highest shell index $C_s(max)$ and is called nucleus of the network.

This algorithm works by recursively pruning the nodes from lower degrees to higher degrees. First, we recursively remove all nodes of degree one until there is no node of degree 1. All these nodes are assigned shell index $C_s=1$. In a similar fashion, nodes of degree 2,3,4, ... are pruned step by step. When we remove nodes of degree $k$, if there appears any node of degree less than $k$, it will also be removed in the same step. All these nodes are assigned shell index $k$. Here, a higher shell index represents higher coreness. Vladimir Batagelj et al. proposed an order $O(m)$ algorithm to calculate the coreness of all nodes, where $m$ is the total number of edges in the graph \cite{batagelj2011fast}.

\subsection{Extensions}


Initially, the k-shell decomposition method was defined for unweighted undirected networks, but recently it has been extended to different types of networks. Garas et al. extended the k-shell method to identify core-periphery structure in weighted networks \cite{garas2012k}. They define the weighted degree that considers both the degree as well as the weights of the connected edges. Then the weighted degree is used while applying the k-shell decomposition method. Eidsaa and Almaas also proposed a method to identify core-periphery structure in weighted networks where they only consider the strength of the nodes while pruning them in each iteration \cite{eidsaa2013s}, and this method is referred to as S-shell or strength decomposition algorithm. The strength of a node is defined as, $s_i=\sum_{j \epsilon \Gamma (i)} W_{ij}$, where $W_{ij}$ denotes the weight of an edge connecting nodes $i$ and $j$. Wei et al. proposed an edge-weighting k-shell method where they consider both the degree as well as the edge-weights and the edge weight is computed by adding the degree of its two end points \cite{wei2015weighted}. The importance of both of these parameters can be set using a tuning parameter, which varies from 0 to 1. If it is set to 0, then the complete importance is given to edge-weights, and if it is set to 1, then the complete importance is given to the degree of the node.

Shell-index assigns the same index values to many nodes, which actually might have different influential power \cite{zareie2018hierarchical, wang2016fast, zeng2013ranking}. Zeng et al. modified the k-shell decomposition method and proposed a mixed degree decomposition (MDD) method, which considers both the residual degree and the exhausted degree of the nodes while assigning them index values \cite{zeng2013ranking}. Liu et al. proposed an improved ranking method that considers both the k-shell value of the node and its distance with the highest k-shell value nodes \cite{liu2013ranking}. The proposed method computes the shortest distance of all nodes with the highest k-shell nodes, so it has high computational complexity. Liu et al. showed that some core-like groups are formed in real-world networks, which are not true-core \cite{liu2015improving}. The nodes in these groups are tightly connected with each other but have very few links outside. Based on this observation, authors filtered out redundant links with low diffusion power but support non-pure core groups to be formed and then apply k-shell decomposition methods. The authors show that the coreness computed on this new graph is a better metric of influential power, and it is highly correlated with spreading power computed using the SIR model in the original graph. 

Researchers also have proposed hybrid centrality measures by combining the k-shell with other existing centrality measures. Hou et al. introduced the concept of all-around score to find influential nodes \cite{hou2012identifying}. All around score of a node can be defined as, $Score=\sqrt{\left \| d \right \|^2 + \left \| C_B \right \| ^2 +\left \| k_s \right \|^2 }$, where $d$ is the degree, $C_B$ is the betweenness centrality, and $k_s$ is the shell-index of the node. The degree takes care of local connectivity of the node, betweenness takes care of shortest paths that represent global information, and k-shell represents the position of the node with respect to the center. The total time complexity of the complete process is $O(nm)$, as it depends on the complexity of betweenness centrality that has the highest complexity. Basaras et al. proposed a hybrid centrality measure based on degree and shell-index and showed that it works better than the traditional shell-index \cite{basaras2013detecting}. Bae and Kim proposed a method where the centrality value of a node is computed based on its neighbors' shell-index value; it thus considers both degree and shell-index value of the nodes \cite{bae2014identifying}. The results show that the proposed method outperforms other methods in scale-free networks with community structure. Tatti and Gionis proposed a graph decomposition method that considers both the connectivity as well as the local density while the k-shell decomposition method only considers the connectivity of the nodes \cite{tatti2015density}. The running time of the proposed algorithm is $O(|V|^2|E|)$. They further proposed a linear-time algorithm that computes a factor-2 approximation of the optimal decomposition value. All the proposed centrality measures have better monotonicity, but all these measures require global information of the network to be computed, and so, they are not favorable in large-scale networks. Basaras et al. showed that the hybrid of degree and coreness (k-shell index) centrality could be efficiently used to identify influential spreaders \cite{basaras2013detecting}. 

\subsection{Approximation Methods}

Lu et al. showed the relationship between degree, H-index \cite{hirsch2005index}, and coreness of a node \cite{lu2016h}. In real world networks, it is observed that the coreness is highly correlated with H-index. H-index family of a node is represented as $H(u) = (h_u^{(0)}, h_u^{(1)}, .... , h_u^{(l)})$, where $l$ is the distance of the farthest node from $u$. $h_u^{(0)}$ is the zero order h-index of the node that is equal to the degree of node $u$, $h_u^{(0)} = k_u$. $h_u^{(i)}$ index of a node is calculated using $h_v^{(i-1)}$ index of its neighbors, where $v \epsilon \Gamma(u)$. If we calculate H-index family of a node then it converges to coreness of the node. This method provides us a new perspective to understand the coreness of a node. In k-shell decomposition method we compute coreness using recursive removal of the nodes but in this method, coreness is computed using an iterative procedure. Fred et al. derived a function to compute the coreness (kshell no) using h-index of the node \cite{fred2017empirical}. The function is proposed as,
\begin{center}
$\frac{m \cdot ln(c(d))}{ln(d)}=\frac{ln(h)}{ln(N)}$
\end{center}
where, $N$ represents total number of nodes, $d$ is degree of node, $m= 1/(\alpha \cdot \beta)$, $\alpha$ is the power-law coefficient for the degree distribution, and $\beta$ is power law exponent from the degree-coreness correlation ($c(d)=d^\beta$). They also empirically verify this correlation on the co-keyword network of mathematical journals.

\subsection{Update in Dynamic Networks}


Real-world networks are highly dynamic, and it will not be feasible to recompute the shell-index of each node for every single change in the network. Li et al. proposed a method to update the shell-index value of the affected nodes whenever a new edge is added or deleted from the network \cite{li2014efficient}. The proposed method updates the coreness of the affected nodes whenever a new edge is added or deleted from the network. Dasari et al. proposed a k-shell decomposition algorithm called ParK that reduces the number of random access calls and the size of the working set while computing the shell-index in larger graphs \cite{dasari2014park}. They further proposed a faster algorithm that involves parallel execution to compute the k-shell in larger graphs. Sariyuce et al. proposed the first incremental k-core decomposition algorithms for streaming networks \cite{sariyuce2013streaming}. They show that the proposed method has a million times speed-up than the original k-shell method on a network having 16 million nodes. Miorandi et al. \cite{miorandi2010k} also proposed methods to rank nodes based on coreness in real-world dynamic networks.

Jakma et al. proposed the first continuous, distributed method to compute shell-index in dynamic graphs \cite{jakma2012distributed}. Pechlivanidou et al. proposed a distributed algorithm based on MapReduce to compute the k-shell of the graph \cite{pechlivanidou2014mapreduce}. Montresor et al. proposed an algorithm to compute k-shell in live distributed systems \cite{montresor2013distributed}. They further show that the execution time of the algorithm is bounded by $1+\sum_{u \in V}[d(u)-k_s(u)]$, and it gives an 80 percent reduction in execution time on the considered real-world networks.

\subsection{Identify Top-k Nodes}

There is not much work on identifying the top-k core nodes in real-world networks or coreness rank. The solutions to these problems will really help network scientists identify influential nodes and better understand the phenomenon of dynamic processes, such as an epidemic, influential spread, or information diffusion taking place on complex networks. Saxena and Iyengar \cite{saxena2018k} proposed a method to estimate the shell-index of a node using local neighborhood information, and the efficiency of the estimator was verified using the monotonicity and SIR spreading model. The authors further discussed hill-climbing based methods to identify the top-ranked nodes using the proposed estimator. The results on real-world networks show that, on average, a top-ranked node can be reached in a small number of steps. The authors also proposed a heuristic method to fast estimate the percentile rank of a node based on the proposed estimator and structural properties of real-world networks.

\subsection{Applications}


In 2010, Kitsak et al. \cite{kitsak2010identification} showed that the nodes of the nucleus are highly influential. If the infection is started from any single node of the core, it will spread more than if it is started from any periphery node. Saxena et al. showed the importance of core nodes in information diffusion on the networks having mesoscale structures \cite{saxena2015understanding}. Several other works support the fact that core-nodes play a crucial role in making information viral \cite{zhao2015k, gupta2016modeling, wu2016parallel, yang2018identifying, gupta2019modeling}. The k-shell method has also been used to identify influential leaders in the communities based on their influence propagation \cite{larsen2017identifying, deborah2019quantify, feng2018identification, saxena2019twitter}.

Catini et al. used shell-index to identify clusters in PubMed scientific publications \cite{catini2015identifying}. To identify the clusters, a graph is created where the nodes are the publications. There is an edge between two nodes if the distance between the corresponding researchers' locations is less than the threshold. In the experiments, the authors have taken the threshold of 1 km. Based on the k-shell decomposition, the authors categorize the cities into monocore and multicore. Later on, the journal impact factors are used to quantify the quality of research of each core. Results show that the k-shell decomposition method can be used to identify the research hub clusters.

Core-periphery structure has been studied in a wide variety of networks, such as financial networks \cite{fricke2015core, barucca2016disentangling}, human-brain networks \cite{bassett2013task, park2013structural}, nervous system of C. elegans worm \cite{chatterjee2007understanding}, blog-networks \cite{obradovic2009journey}, collaboration network \cite{leydesdorff2013international, hu2008visual}, protein interaction networks \cite{luo2009core}, communication network of software development team \cite{crowston2006core, amrit2010exploring, setia2012peripheral, cataldo2008communication}, hollywood collaboration network \cite{cattani2014insiders}, language network \cite{fedorenko2014reworking}, youtube social interaction network \cite{paolillo2008structure}, metabolic networks \cite{zhao2007modular} etc. Carmi et al. used k-shell decomposition method to analyze the hierarchical structure of the network \cite{carmi2007model}. Researchers have studied the evolution of core-periphery structure using k-shell method and proposed evolving models to generate synthetic networks based on their observations \cite{saxena2016evolving, jia2019random, adeniji2017generative}. Karwa et al. proposed a method to generate all graphs for a given shell-index sequence \cite{karwa2017statistical}.

\section{Other Centrality Measures}

We have discussed all main centrality measures defined in network science. Researchers have combined some of these centrality measures or extended them to define new centrality measures based on the requirement. In this section, we will discuss some of these centrality measures with their specific properties.

\begin{enumerate}
\item All-around Score: Identification of the most influential nodes in complex networks is an important issue for more efficient spread of the information. Hou et al. introduced the concept of all-around score to find influential nodes \cite{hou2012identifying}. All around score of a node can be defined as,

\begin{center}
$d=\sqrt{\left \| k \right \|^2 + \left \| C_B \right \| ^2 +\left \| C_S \right \|^2 }$
\end{center}

where $k$ is the degree, $C_B$ is the betweenness centrality, and $C_S$ is the k-shell index of the node. Thus, we consider three metrics to define the importance of a node. The degree takes care of local connectivity of the node, betweenness takes care of shortest paths that represent global information, and k-shell represents the position of the node with respect to the center. The total time complexity of the complete process is $O(nm)$, as it depends on the complexity of betweenness centrality that has the highest complexity. Results show that the all-around distance could be a more effective and stable indicator to show the influential power of a node.

\item Alpha Centrality: Eigenvector centrality does not give good results in some specific kind of networks. Bonacich et al. proposed a centrality measure called alpha centrality that gives similar results as eigenvector centrality \cite{bonacich1972factoring, bonacich2001eigenvector}. It gives good comparable results for the networks where eigenvector centrality can not be applied. Alpha centrality can be defined as,
\begin{center}
$x=\alpha A^T x + e$
\end{center}

where $e$ is a vector having extra information. Parameter $\alpha$ is used to represent the relative importance of endogenous versus exogenous factors. The matrix solution can be written as,
\begin{center}
$x=(I - \alpha A^T)^{-1}e$
\end{center}

Newman et al. showed that under some conditions, the efficacy of eigenvector centrality is impaired, as it gives more importance to a small number of nodes in the network \cite{martin2014localization}. This mainly happens in the networks having hubs or power-law degree distribution. So, they proposed an alternative centrality measure based on the nonbacktracking matrix that gives similar results in dense networks. However, it gives better results where the eigenvector centrality is failed. The complexity of the new centrality measure is the same as the standard one, so it can be easily used for large real-world networks. 

\item Synthesize Centrality ($SC(u)$): Liu et al. defined synthesize centrality to identify opinion leaders in social networks \cite{liu2013identifying}. Opinion leaders have an important influence on information propagation, so it is important to efficiently identify them to understand this dynamic phenomenon. Synthesize centrality is defined as follows:
\begin{center}
$SC(u)= \frac{C_D(u) + C_B(u)}{C_C(u)}$
\end{center}
Experimental results show that if a node is identified as an opinion leader using SC centrality, then there is a high probability that it will be an opinion leader using HITS and PageRank. The proposed method has high efficiency, and its accuracy is increased as the number of opinion leaders increase.

\item C-Index: Yan et al. studied the competence of researchers in a weighted collaboration network and proposed a centrality measure called C-index based on that \cite{yan2013c}. The c-index is computed using the degree, strength, and centrality information of the neighbors of the node. They show that C-index is highly correlated with the competence of the collaboration network. It follows power-law distribution in the weighted scale-free networks. They also propose two more extensions of the c-index centrality measure called iterative c-index and cg-index.

\item Sociability Centrality Index: This centrality metric is used to measure the social skill of a node in a large scale social network \cite{kermani2016introducing}. The proposed centrality measure is based on TOPSIS and Genetic Algorithm. All other centrality measures that we have discussed measure the importance of the node based on its topological location in the network. But the social importance of the node depends on some other features also. The proposed metric considers the psychological and sociological features with the topological location to socially rank a node. The proposed centrality measure is tested on real-world datasets, and it outperforms other existing measures to rank nodes based on social skills.

\end{enumerate}




\section{Centrality Applications in Real-World Networks}

In this section, we are going to discuss applications of centrality measures to understand specific types of networks and the centrality measures that can be applied to the same.

\subsection{Air Transportation Network}

Wang et al. studied the air transportation network of China from the perspective of network science \cite{wang2011exploring}. The authors study the network structure and the centrality measures of each city in the network. They analyze the clustering coefficient, degree centrality, closeness centrality, and betweenness centrality of each node to examine its importance in the network under different contexts. They further study the correlation of these centrality measures with other characteristics of the nodes like the number of seats, frequency of transportation, gross regional domestic product (GRDP), and so on. The network analysis of transportation network can be used to better understand the structure and connectivity of the cities with the main central cities. There are various points that can be looked deeper like how these networks are evolved, how a new city starts making connections in the network, how the location and size of a city affect its connectivity, and so on.

\subsection{Biological Networks}

Erciyes studied a biological network where the nodes represent cells, and edges represent the interactions between the connected cells \cite{erciyes2015analysis}. He performed the centrality analysis on this network to identify important nodes and edges. Other related works include \cite{koschutzki2008centrality, ozgur2008identifying, estrada2010generalized, ghasemi2014centrality, koschutzki2004comparison}; please refer to the papers for more details.

\subsection{Brain Network}

Centrality measures are used to study the brain network, and it helps to understand various brain disorders. Sporns et al. used closeness centrality to identify hubs in the brain network \cite{sporns2007identification}. Martino et al. used degree and eigenvector centrality to study attention-deficit/hyperactivity disorder (ADHD) on the brain connectivity network \cite{di2013shared}.

\subsection{Citation Network}

There are various centrality measures that have been defined to rank researchers based on the analysis of the citation network, such as h-index \cite{hirsch2005index}, g-index, and so on. Vitanov studied some of these centrality metrics like h-index \cite{hirsch2005index}, variations of h-index, g-index, $i_n$-index, on citation network for the assessment of researchers \cite{vitanov2016commonly}. He further discusses m-index, p-index, $IQ_p$-index, A-index, and R-index with respect to the success of a researcher. The paper can be referred to for further details.

\subsection{Sexual Network}

Borgatti has used the degree, closeness, betweenness, and eigenvector centrality for analyzing the sexual network \cite{borgatti1995centrality}. He throws light on one important point of the shortest path centrality measures like closeness and betweenness centrality. The author mentions that these centrality measures can be used if the information or disease spread in all directions at a time, and flow through the shortest paths. But these measures can not help if the information flow in one direction at a time like in a sexual network, a node will be in relation with one node only at any given time. In such type of applications, we can use application-specific centrality measures that are the extensions or hybrid of some main centrality measures. In such types of applications, we need to use the centrality measures based on the trail, not on the shortest paths. One similar example is the rumor spread on social networks, as the rumor can also spread using any path, not only the shortest paths, and it can reach a node any number of times. If we know that the information has been flowed using the shortest paths, only then betweenness centrality can be used to measure the importance of a node in such networks.

\subsection{Social Network}

We have already discussed many applications of centrality measures in social networks. Here we further discuss some very specific centrality measures that have been proposed for social networks. Wang et al. proposed a method to measure the node centrality in directed and weighted networks based on the connectivity of the node \cite{wang2013connectionist}. The proposed method is verified to identify the most influential methods and the results show that the proposed method outperforms other existing methods. Centrality measures have also been used to study students' social networks to understand knowledge diffusion, peer support, homophily, teamwork, academic performance, and so on \cite{saxena2019survey}.

\subsection{Urban Street Network}

Porta et al. used centrality measures to understand the networks of urban streets and intersections \cite{porta2010networks}. Crucitti et al. \cite{crucitti2006centrality} study centrality in urban street patterns of different world cities represented as networks in geographical space. The results show that the self-organized cities have scale-free properties as observed in nonspatial networks, while planned cities do not. Other works on centrality applications on urban street networks include \cite{gao2013understanding, akbarzadeh2019role, agryzkov2019centrality, wang2013exploration}.

\section{Quick points}

\begin{enumerate}
\item Centralities based on shortest paths: closeness, betweenness, stress, graph.
\item Following groups of centrality measures are based on similar concepts:
\begin{enumerate}
\item Closeness, Harmonic
\item Betweenness, Load, Stress, Flow
\item EigenVector, Katz, Pagerank
\item Coreness, H-index, C-index
\end{enumerate}

\item Endpoints, proximal betweenness, k-betweenness, length-scaled betweenness, linearly scaled betweenness, edge betweenness, group betweenness, stress centrality, and load centrality are a modification of betweenness centrality.

\item The performance of various centrality measures is verified either using the well-known ranking of the nodes or using spreading models like SI, SIR, SIS, and so on.

\end{enumerate}

%

\section{Conclusion}\label{con}

In this paper, we have discussed various centrality measures that are used to identify important and influential nodes in real-world networks. As we have discussed, degree centrality is the first basic centrality measure that was used to rank nodes based on their degrees. It was later combined with other parameters, such as the clustering coefficient, the degree of neighbors, the age of ties, and so on, to rank nodes by considering the local neighborhood properties. Then, there are some centrality measures, such as the closeness centrality, betweenness centrality, flow centrality, which are based on the concept of shortest paths. These centrality measures are dependent on each other, and their correlation is discussed in the paper. Next, we discussed eigenvector centrality, pagerank, and coreness; these centrality measures assign the importance to a node based on the importance of its neighbors. The applications of different centrality measures are briefly discussed with the reasons why one specific centrality measure is more applicable than others in the given situation, as observed in different research works based on their experiments. The paper also includes various hybrid centrality measures that have been proposed to rank nodes more efficiently. In the last section, we discuss various real-world networks and the centrality measures that have been applied for the analysis of these networks.


\bibliographystyle{unsrt}

\bibliography{mybib}

\end{document}